\documentstyle{article} \begin{document}
\large
\begin{center}
\bf{WIMP STARS AS DARK MATTER IN THE UNIVERSE}\\
\vspace{3mm}
\it K.M.V.Apparao\\
34,Vibha, R.P.Road, Bandra East,\\
Mumbai 400051, India.\\
\rm
\end{center}
\vspace{5mm}
\par
\Large
It is suggested that the dark matter in the Universe may be made of
stars and black holes made up of WIMP matter.
\large
\vspace{3mm}
\par
The existence of "dark matter" in the Universe is by now established$^{1}$.
Brown Dwarfs, Black Holes, neutrinos, weakly interacting massive particles
(WIMP) and other Super Symmetric particles  have been suggested as
candidates for dark matter$^{1}$. So far the nature of the dark matter
is not known. It is generally assumed that if dark matter is made up of 
particles, it exists as diffuse matter. It is likely, however, that primordial 
 dark matter also condenses due to gravitaional force.
Here we consider the possibility that dark matter exists in the form of 
WIMP stars.
\par
Primordial WIMP matter, which is presumably intermixed with baryonic matter,
 should also condense along with baryonic matter
due to gravitational force. Starting with density fluctuations, the intermixed
material should condense to form proto-galaxies,
 When galaxies are formed from the proto-galaxies, in order to condense,
the baryonic matter gets rid of its angular momentum via various
viscosities. The WIMP matter, however will find it hard to get rid of the
angular momentum due to its weak interaction. This may result in the WIMP 
matter getting separated from the baryonic matter, and will stay in the outer
regions of galaxies. The WIMP matter then may further condense into
WIMP stars.
\par
The formation of neutrino stars has been suggested $^{2}$.
 The Chandrasekhar limit for
such stars is given as$^{2}$,
$$
M_{s}=1.016{\times}10^{10}M_{\odot}(\frac{17.2keV}{m_{\nu}c^{2}})^{2}g^{-1/2}
..............(1)
$$
where M$_{\odot}$ is the mass of the sun, m$_{\nu}$ is the mass  of the
neutrino, c is the velocity of light and g the degeneracy factor. This 
corresponds to a Schwarzschild radius of
$$
R_{s}=1.16({\frac{17.2keV}{m_{\nu}c^{2}}})^{2}g^{-1/2}...................(2)
$$
where R$_{s}$ is given in light days. The radius of the neutrino star
is given by,
$$
R_{0}=6.8631({\frac{M_{\odot}}{M_{S}}})^{1/3}g^{-2/3}(\frac{17.2keV}{m_{\nu}
c^{2}})^{8/3}................(3)
$$
where R$_{0}$ is given in light years. The neutrino stars are suggested to
be at the centres of galaxies$^{2}$.  
\par
The above equations should also hold for stars made of the weakly interacting 
 WIMP particles. The mass of a WIMP is pedicted to be in the range of 10-
1000 GeV $^{3,4}$. The mass limit for a WIMP star, if the mass of the WIMP
is taken as 10 GeV, using (3) is M$_{w}$$\sim$3$\times$10$^{-2}$M$_{\odot}$.
The radius of a WIMP star at the mass limit, using (2), is R$_{w}$=8.9
$\times$10$^{3}$ cm, while the Schwarzschild radius is R$_{s}$=2.8$\times$
10$^{3}$ cm. Since R$_{w}$ is greater than R$_{s}$, the star can be stable.
The denisty of such stars is $\rho$$_{w}$$\sim$2$\times$10$^{19}$ gm cm$^{-3}$.
At such densities even with weak interaction cross-sections, WIMPs will
annihilate if they are Majorana particles. Even if they are Dirac particles,
WIMPs and anti-WIMPs cannot coexist at such high densities. One has to
assume that they are made up of only one kind of WIMPs, by appealing to
processes similar to those that separated baryonic matter and anti-matter,
or processes which resulted in only one kind of baryonic matter in the 
Universe.
\par
It was mentioned earlier that WIMP matter, while condensing, may find it
 difficult to get rid of its angular momentum. It is therefore possible that
WIMP stars with masses greater than the limit given above may exist as 
ellipsoidal stars or disks. If the mass of the WIMP condensate exceeds
 the limit given above, it may also form a WIMP black hole.
\par
Attempts have been made to detect compact objects that make up the dark matter
in the halo of our galaxy$^{5,6}$. In these attempts, gravitaional lensing
of the stars in the Large Magellanic Cloud by compact objects that make up
 the dark matter in the galactic halo (MACHO) was detected. The conclusion
of these studies is that such objects make up a small fraction of the dark
matter required in the halo. The Large Magellanic Cloud is at a high galactic 
latitude, and if the compact objects like WIMP stars are located closer to
the galactic plane,
 as are normal stars, and in the outer regions (due to angular momentum
considerations suggested earlier), the observations cited above$^{5,6}$
may not detect them. Observations using lensing of supernova in external
galaxies$^{7}$ may be able to detect WIMP stars.
\par
It will be interesting to study accretion of gas on to WIMP stars. If one of
them is captured by an ordinary star to form a close binary, then such 
accretion is possible. It may allow identification of WIMP stars or WIMP
black holes.
\par
In conclusion, dark matter in the Universe may exist in the form of WIMP
stars and WIMP black holes.

\vspace{5mm}
\begin{center}
\it{References}
\end{center}
\rm
1. V. Trimble ,\it{Ann.Rev.Astron.Astrophys.},\rm \bf 25,\rm 425-472, 1987.\\
2. R.D.Vollier, D.Trautmann and Tupper G.B.,\it Phys.Lett.,\rm \bf B306,\rm 79-85,1993.\\
3. D.E.Groom et al., Review of Particle Physics,
 \it Eur.Phys. J.,\rm \bf C15,\rm 1-873, 2000.
4. L.Krauss,\it Quintessence,\rm Basic Books, New York, 2000.\\
5. C.Alcock et al.,\it Ap.J.,\rm \bf 542,\rm 281-307, 2000.\\
6. T.Lasserre et al.,\it Astron.Astrophys.,\rm \bf 355,\rm L39, 2000.\\
7. E.Mortsell, A.Goodbar and L.Bergstrom, \it Ap.J.,\rm \bf 559,\rm 53-58, 2001.\\
\end{document}